\documentclass[graybox]{svmult}

\usepackage{type1cm}        
\usepackage{makeidx}         
\usepackage{graphicx}        
\usepackage{multicol}        
\usepackage[bottom]{footmisc}

\usepackage{newtxtext}       %
\usepackage{newtxmath}       
\usepackage{bm}
\usepackage{graphicx} 
\graphicspath{ {Figures/} }

\makeindex             

\newcommand{\mc}{\mathcal}

\begin{document}

\title*{Linear Dynamics \& Control of Brain Networks}
\author{Jason Z. Kim and Danielle S. Bassett}
\institute{Jason Z. Kim \at Department of Bioengineering, University of Pennsylvania, Philadelphia, PA 19104, USA\\ \email{jinsu1@seas.upenn.edu}
\and Danielle S. Bassett \at Departments of Bioengineering, Electrical \& Systems Engineering, Physics \& Astronomy, Neurology, \& Psychiatry, University of Pennsylvania, Philadelphia, PA 19104, USA\\ \email{dsb@seas.upenn.edu}}
%
%
\maketitle

\abstract{The brain is an intricately structured organ responsible for the rich emergent dynamics that support the complex cognitive functions we enjoy as humans. With around $10^{11}$ neurons and $10^{15}$ synapses, understanding how the human brain works has proven to be a daunting endeavor, requiring concerted collaboration across traditional disciplinary boundaries. In some cases, that collaboration has occurred between experimentalists and technicians, who offer new physical tools to measure and manipulate neural function. In other contexts, that collaboration has occurred between experimentalists and theorists, who offer new conceptual tools to explain existing data and inform new directions for empirical research. In this chapter, we offer an example of the latter. Specifically, we focus on the simple but powerful framework of linear systems theory as a useful tool both for capturing biophysically relevant parameters of neural activity and connectivity, and for analytical and numerical study. We begin with a brief overview of state-space representations and linearization of neural models for non-linear dynamical systems. We then derive core concepts in the theory of linear systems such as the impulse and controlled responses to external stimuli, achieving desired state transitions, controllability, and minimum energy control. Afterwards, we discuss recent advances in the application of linear systems theory to structural and functional brain data across multiple spatial and temporal scales, along with methodological considerations and limitations. We close with a brief discussion of open frontiers and our vision for the future.}

\section{Emergence in the Structure and Function of Complex Systems}
\label{sec:1} 

In the observable world, some of the most beautiful and most puzzling phenomena arise in physical and biological systems characterized by heterogeneous interactions between constituent elements. For example in materials physics, heterogeneous interactions between particles in granular matter (such as a sand pile) constrain whether the matter acts as a liquid (flowing with gravity) or a solid (supporting load-bearing) \cite{maier2017emergence,kivelson2016defining}. In sociology, heterogeneous interactions between humans in a society are thought to be responsible for surges in online activity, peaks in book sales, traffic jams, and correlated spikes in demand for emergency services \cite{lynn2018surges}. In biology, heterogeneous interactions between computational units in the brain are thought to support a divergence of the correlation length, an anomalous scaling of correlation fluctuations, and the manifestation of mesoscale structure in patterns of functional coupling between units, all features that allow for a diversity of dynamics underlying a diversity of cognitive functions \cite{bassett2011understanding,haimovici2013brain}. The feature of these systems that often drives our fascination is the capacity for heterogeneous interactions to produce suprising dynamics, in the form of drastic state transitions, spikes of collective activity, and multiple accessible dynamical regimes.

Because element-element interactions are heterogeneous in such systems, traditional approaches from statistical mechanics -- such as continuum models and mean-field approximations -- fail to offer satisfying explanations for system function. There exists a critical need to develop alternative approaches to understand how interactions map to emergent behavior. The need is particularly salient in the context of neural systems, where such an understanding could directly inform models of neurological disease and psychiatric disorders \cite{braun2018maps,stam2014modern}. Moreover, gaining such an understanding is a prerequisite for the well-reasoned development of interventions \cite{tang2018control}, whether in the form of brain stimulation \cite{downar2014anhedonia,medaglia2018network}, pharmacological agents \cite{gass2018antagonism,braun2016dynamic}, or other therapies \cite{yang2018network}. Technically, such interventions in systems characterized by heterogeneous interactions can be parsimoniously considered as forms of network control, thus motivating extensive recent interest in the utility of network control theory for neural systems \cite{tang2018control}. 
 
Despite the generic importance of understanding how interactions map to emergent properties, and the specific importance of understanding that mapping in the human brain, progress towards that understanding has remained surprisingly slow. Some efforts have sought to develop detailed multiscale computational models \cite{markram2015reconstruction}. Yet such efforts are faced with the ever-present quandary that, in point of fact, ``The best material model of a cat is another, or preferably the same, cat'' \cite{rosenblueth1945role}. Detailed models are difficult to construct, intractable to analytic approaches, require extensive time to simulate, contain parameters that are frequently underconstrained by experimental data, and in the end produce dynamics that are themselves difficult to understand or to explain from any specific choices in the model. In contrast, approaches from physics consider natural phenomena as if dynamics at macroscopic length scales were almost independent of the underlying, shorter length scale details \cite{machta2013parameter}. A hallmark of effective physical theories is a marked compression of the full parameter space into a few governing variables that are sufficient to describe the observables of interest at the scale of interest. Interestingly, recent theoretical work demonstrates that such simple models are the natural culmination of processes maximizing the information learned from finite data \cite{mattingly2018maximizing}.

Here we embrace simplicity by considering the utility of linear systems theory for the understanding and control of neural systems comprised of computational units coupled by heterogeneous interactions. We begin by placing our remarks within the context of quantitative dynamical models of neurons and their interactions, as well as the spatial and temporal considerations inherent in choosing such models. We will then turn to a discussion of approximations to those dynamical models, the incorporation of exogeneous control input, and model linearization. Our treatment then naturally brings us to a discussion of the theory of linear systems, as well as their response to perturbative impulses, and to explicit control strategies. We lay out the formalism for probing state transitions, controllabilty, and the minumum control energy needed for a given state transition. After completing our formal treatment, we discuss the application of linear systems theory to neural systems, and efforts to map network architecture to control properties. We close with a description of several particularly pertinent methodological considerations and limitations, before outlining emerging frontiers.

\section{Quantitative Dynamical Models of Neural Systems \& Interactions}
\label{sec:2} 
\begin{figure}[h!]
	\centering
	\includegraphics[width=1.0\columnwidth]{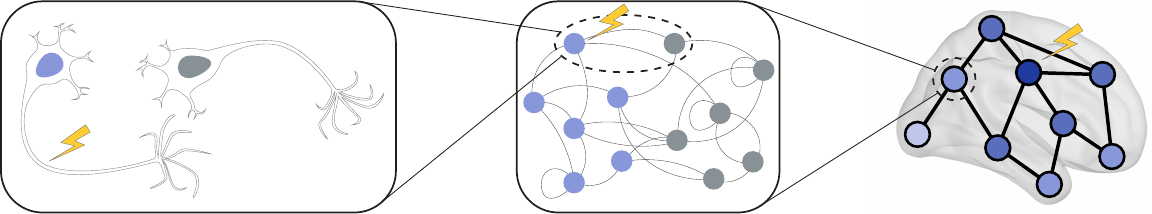}
	\caption{\textbf{Schematic of neural models and controlling perturbations at different scales.} Here, the Hodgkin-Huxley model describes the biophysical behavior of single neurons (\emph{left}) that may be excitatory (blue) or inhibitory (gray). The artificial neuron models describe the simplified weighted connections and binary states of many neurons (\emph{center}). The Wilson-Cowan model describes the activity of large neural populations in a region (\emph{right}) or in a cortical column by modeling the excitatory and inhibitory connections of each population. In each case, a controlling perturbation (yellow) can affect the neural system at different scales.}
	\label{fig:scales}
\end{figure}

Historically, many neural behaviors and mechanisms have been successfully modeled quantitatively. Here we briefly describe several illustrative examples of such models. The classic fundamental biophysical model of a single neuron (Fig.~\ref{fig:scales}, left) was developed by Alan Hodgkin and Andrew Huxley in 1952 (see \cite{hodgkinhuxley} for details). The model is now known as the \emph{Hodgkin-Huxley} model. It treats a segment of a neuron as an electrical circuit, where the membrane (capacitor) and voltage-gated ion channels (resistors) are parallel circuit elements. The time-evolution of membrane voltage, $V_m$, between the inside and the outside of the neuron is given by
\begin{align*}
C_m \dot{V_m}(t) &= \bar{g}_Kn^4(t)(V_K-V_m) + \bar{g}_{Na}m^3(t)h(t)(V_{Na}-V_m) + \bar{g}_l(V_l-V_m) + I(t),
\end{align*}
where $C_m$ is the membrane capacitance, $\bar{g}_K, \bar{g}_{Na}, \bar{g}_l$ are maximum ion conductances for potassium, sodium, and passive leaking ions, and $I$ is an external stimulus current, all per unit area. In addition, $V_K, V_{Na}, V_l$ represent the reversal potential of these ions. The variables $n, m, h$ vary between 0 and 1, and model the ion channel gate kinetics to determine the fraction of open sodium ($m,h$) and potassium ($n$) channels
\begin{align*}
\dot{n}(t) &= \alpha_n(V_m(t))(1-n(t)) - \beta_n(V_m(t))n(t)\\
\dot{m}(t) &= \alpha_m(V_m(t))(1-m(t)) - \beta_m(V_m(t))m(t)\\
\dot{h}(t) &= \alpha_h(V_m(t))(1-h(t)) - \beta_h(V_m(t))h(t),
\end{align*}
where the functions $\alpha_i(V_m)$ and $\beta_i(V_m)$ are empirically determined. These segments are then spatially connected together, such that the propagation of an action potential across a neuron is modeled by a set of partial differential equations. Due to the biophysical realism of variables and parameters, this model can make powerful and accurate predictions of neuron activity in different environments and stimulation regimes \cite{cano2017intermittency,goldwyn2011what,teka2016power}. Simplified versions of this model, such as the FitzHugh-Nagumo model \cite{fitzhugh}, can also produce many of the same neuronal dynamics.

However, many complex behaviors of neural systems arise from \emph{interactions} between multiple neurons. With four variables (membrane voltage, gates) and even more parameters to model the behavior of a single neuron, the space of models to explore interacting neurons quickly becomes intractable to both analytical and numerical interrogation. An alternative approach is to capture the simplest aspects of neural interactions that are crucial for the phenomenon of interest. Such was the approach taken by Warren McCulloch and Walter Pitts \cite{mccullochpitts}, who developed what would later become a canonical model of an artificial neuron. In this model, each neuron $i$ at any point in time $t$ exists in one of two states: firing $x_i(t) = 1$ or not firing $x_i(t) = 0$. The state of the neuron is determined by a weighted sum of inputs from connected neurons at the previous time step. Then, neuron $i$ in a system of $N$ neurons evolves in time as
\begin{align*}
x_i(t+1) = f_i\left(\sum_{j=1}^N w_{ij}x_j(t)\right),
\end{align*}
where $w_{ij}$ is the strength of excitation ($w_{ij}>0$) or inhibition ($w_{ij}<0$) from neuron $j$ to neuron $i$, and function $f_i$ is typically a thresholding function (Fig.~\ref{fig:scales}, center). Instantiations and extensions of this model are used to study associative memory (Hopfield \cite{hopfield}), machine learning (perceptron \cite{rosenblatt}), and cellular automata \cite{hedlund}.

In many cases, the sheer number of neurons and interactions renders even these simple models difficult to study. A typical solution is to instead model the average activity of a \emph{population} of neurons. This is the approach taken by Hugh Wilson and Jack Cowan \cite{wilson} in the \emph{Wilson-Cowan} model. Here, a group of neurons is separated into excitatory and inhibitory populations, where the fraction of cells firing at time $t$ in each population is $E(t)$ and $I(t)$, respectively, that evolve in time as
\begin{align*}
\tau_e \dot{E}(t) &= -E(t) + (k_e-r_eE(t))S_e(c_1E(t) - c_2I(t) + P(t))\\
\tau_i \dot{I}(t) &= -I(t) + (k_i-r_iI(t))S_i(c_3E(t) - c_4I(t) + Q(t)).
\end{align*}
Here, $c_1, c_2 > 0$ represent connection strength into the excitatory population, and $c_3, c_4 > 0$ represent connection strength into the inhibitory population, $r_e, r_i$ are the refractory periods, and $S_e, S_i$ are sigmoid functions from the distribution of neuron input thresholds for firing. Such models produce oscillations such as those observed in non-invasive measurements of large-scale brain activity (Fig.~\ref{fig:scales}, right) in patients with epilepsy \cite{shusterman}.

In these and many other models, a common theme is the tradeoff between realism and tractability. We desire sufficient realism to study crucial features of neural systems such as the activity of each unit, the interaction strength between units, the connection topology, and the effect of external stimulation. We also desire sufficient tractability (either to analytical or numerical interrogation) to make consistent and meaningful predictions about our neural system by understanding relations between the model parameters and the model behavior. In this chapter, we will discuss one such model from the theory of linear dynamical systems.

\subsection{Spatial and Temporal Considerations}
\label{subsec:1} 
When modeling neural systems, an immediately salient consideration is the vast range of spatial and temporal scales at which nontrivial -- and thus quite interesting -- dynamics occur. It stands to reason that the most relevant type of model for understanding a given phenomenon depends on the spatiotemporal scale at which that phenomenon is observed. For example, consider the fact that while it is generally known that certain sensory regions such as the visual cortex are both anatomically linked to and functionally responsible for sensory inputs, it is more difficult to assign a set of neurons that are necessary for distributed cognitive processes such as attention and cognitive control. Thus, biophysical models at the level of single neurons may be viable for simulating receptive fields in visual processing, but may be less useful for studies of task-switching or gating. Similarly, consider the fact that a single neuron may fire every few milliseconds, while human reaction times are on the order of hundreds of milliseconds, and brain-wide fluctuations in activity on the order of seconds. Thus, the form of the model considered should match the temporal scales of the behavior to be studied. 

From a modeling perspective, balancing these considerations of spatial and temporal scales with model realism impacts the category of model that has the greatest utility. If one wishes to consider small spatial scales, then a rather simplistic neuron-level model such as the McCulloch-Pitts may be particularly useful, where each neural unit has \emph{discrete states} such that each neuron $i$ is either firing $x_i(t) = 1$ or not $x_i(t) = 0$. In contrast, if one wishes to consider larger spatial scales characteristic of distributed cognitive processes, it may be more appropriate to consider models in which each neural unit reflects the average population activity of a brain region as a \emph{continuous state}, where $x_i(t)$ is a real number. Similar considerations are relevant and important in the time domain. For models that assume fairly uniform delays in neuronal interactions such as the McCulloch-Pitts, a \emph{discrete time} model where time evolves in integer increments may be appropriate. In contrast, if the timing of interactions between neural units such as myelinated \emph{versus} unmyelinated axons is heterogeneous, a \emph{continuous time} model may be more suitable, where time $t$ is a real number. 

In addition to affecting the definition of neural activity and the nature of its propagation, these considerations also affect the meaning of interactions between units. In a neuron-level model whose units reflect neurons, the unit-to-unit interactions may represent structural synapses between neurons. In contrast, in a population model whose units reflect average neural activity of a brain region, unit-to-unit interactions may represent a summary measure of the collective strength or extent of structural connections between regions. Both types of connections can be empirically measured using either invasive (staining, flourescence imaging, tract tracing \cite{oh}) and non-invasive (tractography \cite{basser}) methods. The specific type of interaction studied constrains the sorts of inferences that one can draw from the subsequent model, as well as the types of model-generated hypotheses that one can test in new experiments.

\subsection{Dynamical Model Approximations}
\label{subsec:2} 
\begin{figure}[h!]
	\centering
	\includegraphics[width=0.8\columnwidth]{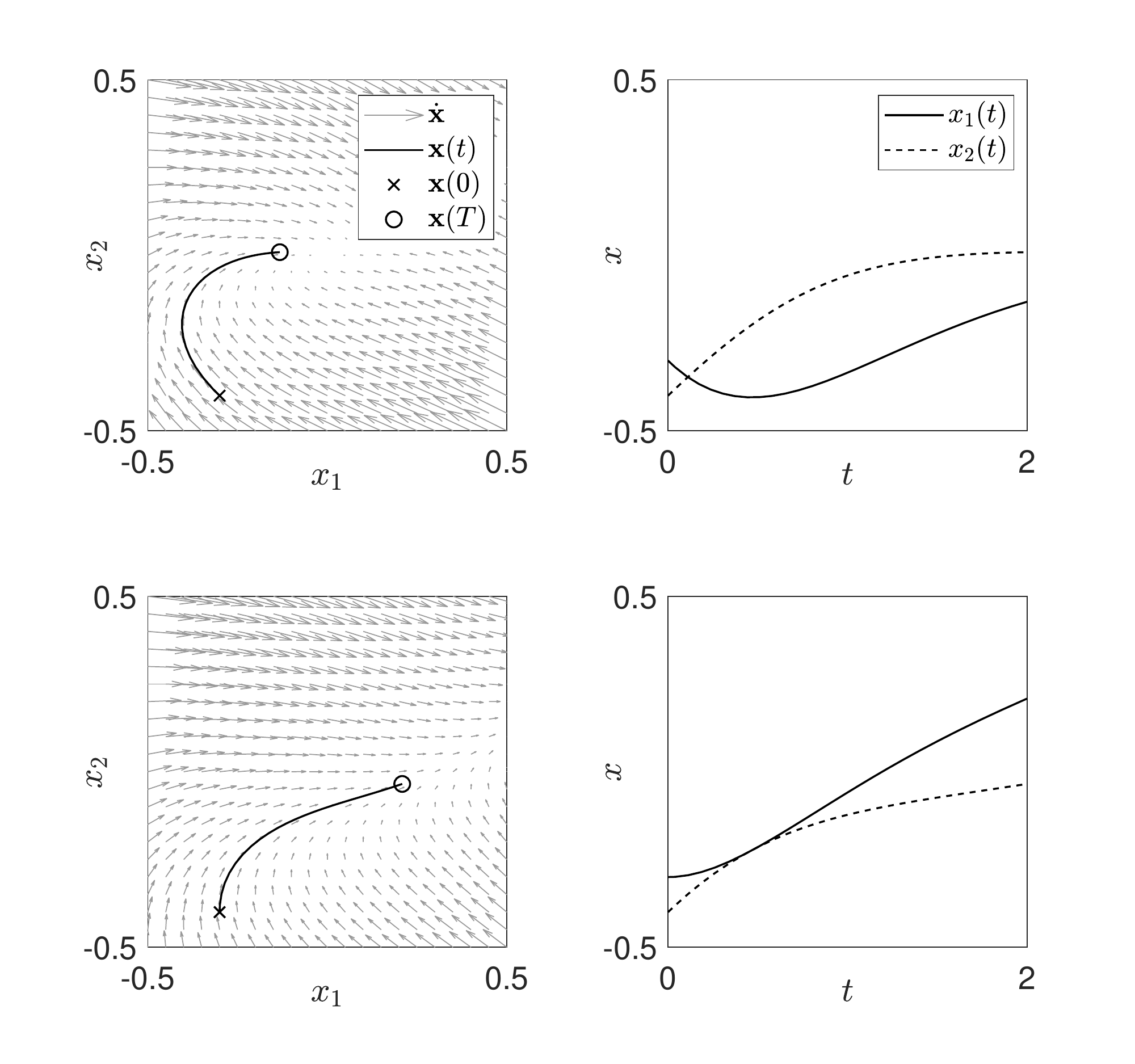}
	\caption{\textbf{Vector fields and trajectories, with and without control inputs.} Example simple vector field of two states with a particular trajectory from initial condition $\bm{x}(0) = [-0.3; -0.4]$ (\emph{top left}) in state space, with the corresponding plot of each state over time (\emph{top right}), and the corresponding vector field and trajectory with control input $u(t) = 0.5$ (\emph{bottom left}) with corresponding states over time (\emph{bottom right}).}
	\label{fig:example}
\end{figure}

Both here and in the following sections, we will consider systems with both continuous state and time. However, we note that the theory of linear systems extends naturally to discrete time systems as well. We begin our formulation with a set of $N$ neural units, where each unit has an associated level of activity $x_i(t)$ that is a real number at some time $t \geq 0$ that is also a real number. Then the collection of activity for all units into column vector $\bm{x}(t) = [x_1(t); x_2(t); \dotsm; x_N(t)]$ is called the \emph{state} of our system at time $t$. For example, in the Hodgkin-Huxley equations, our state vector is $\bm{x} = [V; n; m; h]$. In many models including Hodgkin-Huxley, the time evolution of the system states can be written as a vector derivative
\begin{align*}
\underbrace{
\begin{bmatrix}
\dot{x}_1(t)\\
\dot{x}_2(t)\\
\vdots\\
\dot{x}_N(t)
\end{bmatrix}}_{\dot{\bm{x}}(t)}
=
\underbrace{
\begin{bmatrix}
f_1(\bm{x}(t))\\
f_2(\bm{x}(t))\\
\vdots\\
f_N(\bm{x}(t))
\end{bmatrix}}_{\bm{f}(\bm{x}(t))},
\end{align*}
where $\bm{f}$, the vector of functions $f_i$, determines how the system states change, $\dot{\bm{x}}$, at every particular state $\bm{x}$. We can think of these equations as generating a vector field, where at each point $\bm{x}$, we draw an arrow with magnitude and direction equal to $\bm{f}(\bm{x})$. As an example, consider the following two neuron system $x_1, x_2$ that evolves in time as
\begin{align*}
\dot{x}_1(t) &= 2x_2(t) - \sin(x_1(t))\\
\dot{x}_2(t) &= x_1^2(t) - x_2(t),
\end{align*}
where the vector field and example trajectory from initial state $\bm{x}(0) = [-0.3; -0.4]$ is shown (Fig.~\ref{fig:example}, top). Note how at every point $x_1,x_2$, the above equation determines a vector of motion $\dot{\bm{x}}$ that the system traces from the initial point. This quantitative modeling of neural dynamics allows us to study and predict the response of our neural system to changes in interaction strength or external stimulation.

\subsection{Incorporating Exogenous Control}
\label{subsec:3} 
While modeling intrinsic system behavior is already a broad topic of current research, there is an increasing need for the principled study of therapeutic interventions to correct dysfunctional neural activity. These interventions may take the form of targeted invasive (deep bran stimulation) or non-invasive (transcranial magnetic stimulation) inputs, or more diffusive drug treatments. Hence, in our modeling efforts we also often desire to incorporate the effect of some external stimuli $u_1(t), \dotsm, u_k(t)$. We collect these stimuli into a vector $\bm{u}(t) = [u_1(t); u_2(t); \dotsm; u_k(t)]$, and include their effect on the rates of change of system states in our function
\begin{align*}
\underbrace{
	\begin{bmatrix}
	\dot{x}_1(t)\\
	\dot{x}_2(t)\\
	\vdots\\
	\dot{x}_N(t)
	\end{bmatrix}}_{\dot{\bm{x}}(t)}
=
\underbrace{
	\begin{bmatrix}
	f_1(\bm{x}(t),\bm{u}(t))\\
	f_2(\bm{x}(t),\bm{u}(t))\\
	\vdots\\
	f_N(\bm{x}(t),\bm{u}(t))
	\end{bmatrix}}_{\bm{f}(\bm{x}(t),\bm{u}(t))}.
\end{align*}
As an example in our two unit system, we can apply an input to the first unit
\begin{align*}
\dot{x}_1(t) &= 2x_2(t) - \sin(x_1(t)) + u(t)\\
\dot{x}_2(t) &= x_1^2(t) - x_2(t),
\end{align*}
thereby changing our system of equations. We plot the vector field and trajectory of our system under some constant input $u(t) = 0.5$ (Fig.~\ref{fig:example}, bottom). Notice how the control input changes the trajectory and final state of our system by modifying the vector field. Also notice that our input only shifts the $x_1$ component of our vectors because we only stimulate $x_1$. These abilities to map neural interactions $\bm{f}$ to the full trajectory of activity $\bm{x}(t)$, and to find control inputs $\bm{u}(t)$ that drive our neural system to a desired final state $\bm{x}(T)$ are among the core contributions of linear systems theory.

\subsection{Model Linearization}
\label{subsec:4} 
While we have a quantitative framework for the evolution of a controlled neural system, there are no general principles for determining the full trajectory $\bm{x}(t)$ or control input $\bm{u}(t)$ to reach a desired final state for a general nonlinear system. In systems of only a few neural units, there exist several powerful numerical and analytic tools. However, the study and control of large neural systems is made difficult by our inability to know how a stimulus will affect our system without first simulating the full trajectory. Further, for multiple stimuli, the number of possible stimulus patterns grows exponentially.

A special class of simplified systems called \emph{linear systems} circumvents this issue. In our state representation, a linear system is described by
\begin{align}
\label{eq:linear_system}
\underbrace{
	\begin{bmatrix}
	\dot{x}_1(t)\\
	\dot{x}_2(t)\\
	\vdots\\
	\dot{x}_N(t)
	\end{bmatrix}}_{\dot{\bm{x}}(t)}
=
\underbrace{
\begin{bmatrix}
a_{11} & a_{12} & \dotsm & a_{1N}\\
a_{21} & a_{22} & \dotsm & a_{2N}\\
\vdots & \vdots & \ddots & \vdots\\
a_{N1} & a_{N2} & \dotsm & a_{NN}
\end{bmatrix}}_A
\underbrace{
\begin{bmatrix}
x_1(t)\\
x_2(t)\\
\vdots\\
x_N(t)
\end{bmatrix}}_{\bm{x}(t)}
+
\underbrace{
\begin{bmatrix}
b_{11} & b_{12} & \dotsm & b_{1k}\\
b_{21} & b_{22} & \dotsm & b_{2k}\\
\vdots & \vdots & \ddots & \vdots\\
b_{N1} & b_{N2} & \dotsm & b_{Nk}
\end{bmatrix}}_B
\underbrace{
\begin{bmatrix}
u_1(t)\\
u_2(t)\\
\vdots\\
u_k(t)
\end{bmatrix}}_{\bm{u}(t)},
\end{align}
that is characterized by the time evolution of any state $\dot{x}_i(t)$ being a weighted sum of current states $\sum_{j=1}^N a_{ij}x_j(t)$ and external inputs $\sum_{j=1}^k b_{ij}u_j(t)$. We see that our two-unit system is \emph{not} linear, because the first state $\dot{x}_1(t)$ depends on $\sin(x_1(t))$, and the second state $\dot{x}_2(t)$ depends on $x_1^2(t)$, and is therefore a \emph{non-linear} system. 

\begin{figure}[h!]
	\centering
	\includegraphics[width=0.8\columnwidth]{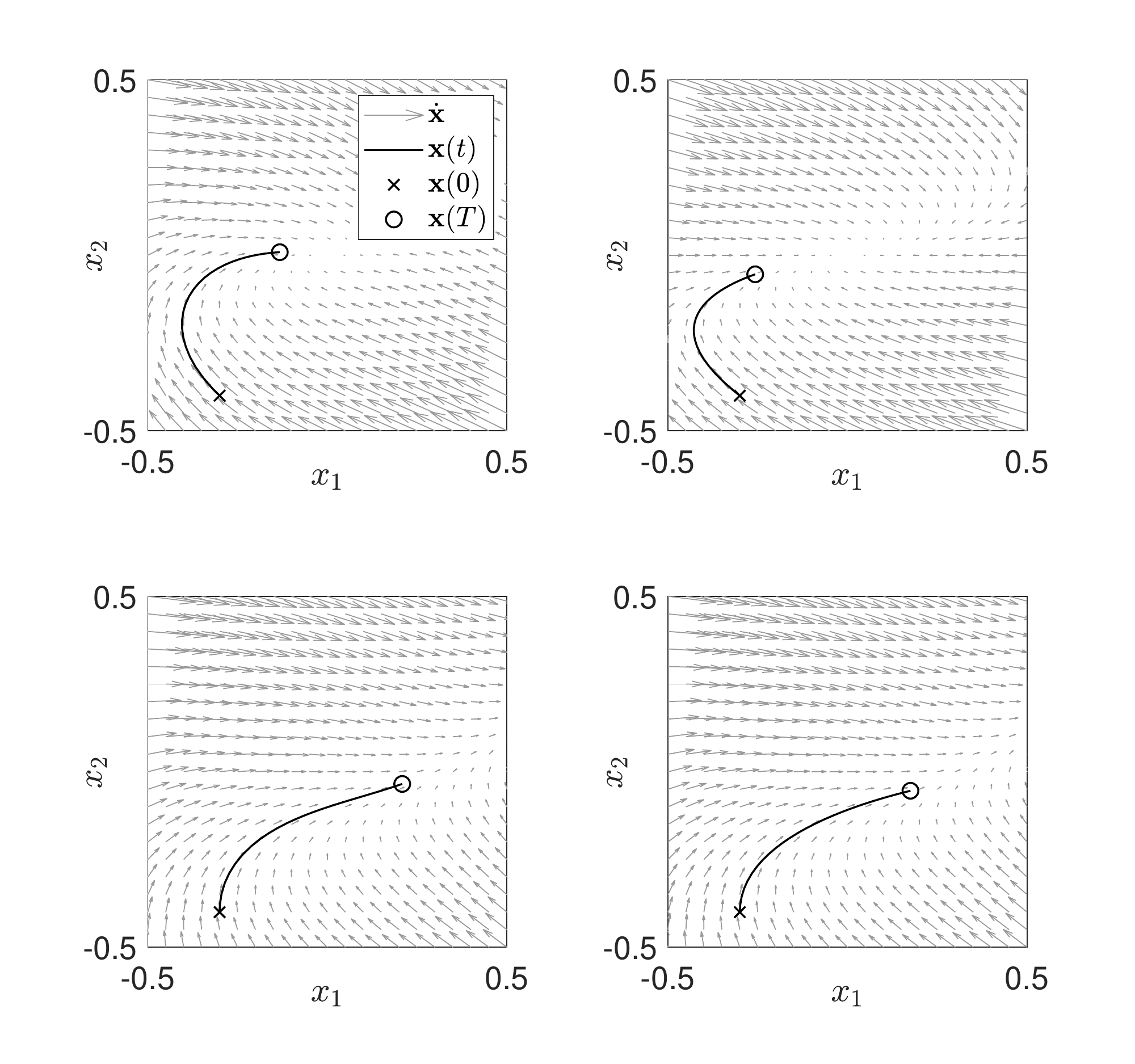}
	\caption{\textbf{Vector fields and trajectories for a nonlinear system and its linearized form.} Example vector field of two states with a particular trajectory from initial condition $\bm{x}(0) = [-0.3; -0.4]$ for the uncontrolled nonlinear system (\emph{top left}), the uncontrolled linear system (\emph{top right}), the controlled nonlinear system (\emph{bottom left}) and the controlled linear system (\emph{bottom right}).}
	\label{fig:example_linear}
\end{figure}

To transform the nonlinear system $\dot{\bm{x}} = \bm{f}(\bm{x},\bm{u})$, into a linear system $\dot{\bm{x}} = A\bm{x} + B\bm{u}$, we can create an approximate model of our vector field about a particular operating state $\bm{x}^*$ and input $\bm{u}^*$. We first evaluate the dynamics at this operating point, $\bm{f}(\bm{x}^*,\bm{u}^*)$. Then we approximate the vector field along small deviations from this point by computing the derivative of $\bm{f}(\bm{x},\bm{u})$ with respect to the states to get matrix $A$, and with respect to control inputs to get matrix $B$
\begin{align*}
A = 
\left.
\begin{bmatrix}
\frac{\partial f_1}{\partial x_1} & \frac{\partial f_1}{\partial x_2} & \dotsm & \frac{\partial f_1}{\partial x_N}\\
\frac{\partial f_2}{\partial x_1} & \frac{\partial f_2}{\partial x_2} & \dotsm & \frac{\partial f_2}{\partial x_N}\\
\vdots & \vdots & \ddots & \vdots\\
\frac{\partial f_N}{\partial x_1} & \frac{\partial f_N}{\partial x_2} & \dotsm & \frac{\partial f_N}{\partial x_N}
\end{bmatrix}
\right\vert_{\bm{x} = \bm{x}^*,\bm{u} = \bm{u}^*}
\hspace{1cm}
B = 
\left.
\begin{bmatrix}
\frac{\partial f_1}{\partial u_1} & \frac{\partial f_1}{\partial u_2} & \dotsm & \frac{\partial f_1}{\partial u_k}\\
\frac{\partial f_2}{\partial u_1} & \frac{\partial f_2}{\partial u_2} & \dotsm & \frac{\partial f_2}{\partial u_k}\\
\vdots & \vdots & \ddots & \vdots\\
\frac{\partial f_N}{\partial u_1} & \frac{\partial f_N}{\partial u_2} & \dotsm & \frac{\partial f_N}{\partial u_k}
\end{bmatrix}
\right\vert_{\bm{x} = \bm{x}^*,\bm{u} = \bm{u}^*}.
\end{align*}
Then, for states near $\bm{x}^*$ and inputs near $\bm{u}^*$, the vector field is approximately
\begin{align}
\label{eq:linearize}
\dot{\bm{x}}(t) = \bm{f}(\bm{x},\bm{u}) \approx \bm{f}(\bm{x}^*,\bm{u}^*) + A(\bm{x}(t) - \bm{x}^*) + B(\bm{u}(t) - \bm{u}^*).
\end{align}
A typical operating point for the input is $\bm{u}^* = \bm{0}$ corresponding to no input, because neural stimulation is viewed as a perturbation to the natural and unstimulated dynamics. A typical operating point for the state $\bm{x}^*$ is a \emph{fixed point} where $\bm{f}(\bm{x}^*,\bm{u}^*) = \bm{0}$, because then the evolution of our system Eq.~\ref{eq:linearize} only depends on deviations from the point, and not on its actual value. Finally, we can write the linearized equation explicitly as a function of these deviations through a change of variables $\bm{y}(t) = \bm{x}(t) - \bm{x}^*$, 
\begin{align*}
\dot{\bm{y}}(t) = \dot{\bm{x}}(t) \approx A\bm{y}(t) + B\bm{u}(t).
\end{align*}
We will continue to use variable $\bm{x}$ instead of $\bm{y}$ with the understanding that it represents deviations from the fixed point. For example, in our two-unit system, we can linearize about $x_1^* = 0, x_2^* = 0, u^* = 0$ to yield
\begin{align*}
\begin{bmatrix}
\dot{x}_1(t)\\
\dot{x}_2(t)
\end{bmatrix}
\approx
\begin{bmatrix}
-1 &  2\\
 0 & -1
\end{bmatrix}
\begin{bmatrix}
x_1(t)\\
x_2(t)
\end{bmatrix}
+
\begin{bmatrix}
1\\
0
\end{bmatrix}
u(t).
\end{align*}
We show the vector fields and trajectories for both the nonlinear and linear equations without control where $u(t) = 0$ (Fig.~\ref{fig:example_linear}, top), and with control where $u(t) = 0.5$ (Fig.~\ref{fig:example_linear}, bottom) from the same initial condition, and we notice that in the neighborhood of $x_1^* = 0, x_2^* = 0$, the field and trajectories are similar. Hence, by linearizing our neural dynamics about $\bm{x}^*,\bm{u}^*$, we can preserve the behavior of our neural system at state $\bm{x}(t)$ and inputs $\bm{u}(t)$ near this point, while enabling the use of powerful tools developed in the next section. 

\section{Theory of Linear Systems}
\label{sec:3} 
A useful model for therapeutic intervention in a neural system should capture both how the activity over time depends on the connections between neural units, and how to change the activity in a desired way through stimulation. Now that we have a model that captures features of neural activity and connectivity in a linearized form, we will develop equations that yield precisely these features. Specifically, we will first determine the system's response to control through mathematical relations as opposed to simulations. Then we will use these principles to design stimuli that optimally guide our system from some initial state $\bm{x}(0)$ to some final state $\bm{x}(T)$.

\subsection{Impulse Response}
\label{subsec:5} %
First, we find the natural evolution of system states from some initial neural state $\bm{x}(0)$ without any external input. This task amounts to finding the state trajectory $\bm{x}(t)$ that solves our dynamic equation $\dot{\bm{x}}(t) = A\bm{x}(t)$. For scalar systems where $x(t)$ is not a vector, we are reminded of the solution to $\dot{x} = ax$:
\begin{align*}
\frac{dx}{dt} &= ax\\
\frac{1}{x}dx &= adt\\
\int\frac{1}{x}dx &= \int a dt + c\\
\ln|x| &= at+c\\
x &= Ce^{at},
\end{align*}
where the constant is the initial condition $C = x(0)$. We can prove that this solution satisfies $\dot{x} = ax$ by using a Taylor series of the exponential function $e^{at} = \sum_{k=0}^\infty \frac{(at)^k}{k!}$. Taking the time derivative of $x(t) = e^{at}$, we see $\dot{x} = ax$
\begin{align*}
\frac{d}{dt}e^{at} 
&= \frac{d}{dt}\left(1 + \frac{at}{1!} + \frac{a^2t^2}{2!} + \frac{a^3t^2}{3!} + \dotsm + \frac{a^kt^k}{k!} + \dotsm\right)\\
&= 0 + \frac{a}{1!} + 2\frac{a^2t}{2!} + 3\frac{a^3t^2}{3!} + \dotsm + k\frac{a^kt^{k-1}}{k!} + \dotsm\\
&= a\left(1 + \frac{at}{1!} + \frac{a^2t^2}{2!} + \dotsm + \frac{a^kt^k}{k!} + \dotsm\right)\\
&= ae^{at}.
\end{align*}
A \emph{matrix exponential} is defined exactly the same as above with $e^{At} = \sum_{k=0}^\infty \frac{(At)^k}{k!}$, and we again show that the time derivative satisfies the vector relation $\dot{\bm{x}}(t) = A\bm{x}(t)$
\begin{align*}
\frac{d}{dt}e^{At} 
&= \frac{d}{dt}\left(1 + \frac{At}{1!} + \frac{A^2t^2}{2!} + \frac{A^3t^2}{3!} + \dotsm + \frac{A^kt^k}{k!} + \dotsm\right)\\
&= 0 + \frac{A}{1!} + 2\frac{A^2t}{2!} + 3\frac{A^3t^2}{3!} + \dotsm + k\frac{A^kt^{k-1}}{k!} + \dotsm\\
&= A\left(1 + \frac{At}{1!} + \frac{A^2t^2}{2!} + \dotsm + \frac{A^kt^k}{k!} + \dotsm\right)\\
&= Ae^{At}.
\end{align*}
Hence, we see that the following solution
\begin{align}
\label{eq:impulse}
\bm{x}(t) = e^{At}\bm{x}(0),
\end{align} 
satisfies our dynamic equation. Here, the matrix exponential $e^{At}$ is called the \emph{state transition matrix}, and Eq.~\ref{eq:impulse} is called the \emph{impulse response} of our system. Hence, we can find the state at any time $T$ without solving for intermediate states $0 < t < T$.

As an example in our two unit model, to find the state of our system at $T = 2$ given an initial start at $\bm{x}(0) = [-0.3; -0.4]$, we can use software to numerically compute the matrix exponential at time $t = 2$, and multiply by our initial state Eq.~\ref{eq:impulse}
\begin{align*}
\bm{x}(2) 
= 
e^{2A}\bm{x}(0)
=
\begin{bmatrix}
0.1353 & 0.5413\\
0 & 0.1353
\end{bmatrix}
\begin{bmatrix}
-0.3\\ -0.4
\end{bmatrix}
=
\begin{bmatrix}
-0.2571\\
-0.0541
\end{bmatrix},
\end{align*}
which agrees with the simulation results (Fig.~\ref{fig:example_linear}).

\subsection{Control Response}
\label{subsec:6} %
Next, we derive the system response from an initial state $\bm{x}(0)$ to some controlling input $\bm{u}(t)$ through some algebraic manipulation and calculus. We begin with our system equations $\dot{\bm{x}}(t) - A\bm{x}(t) = B\bm{u}(t)$, and multiply both sides by a matrix exponential
\begin{align*}
e^{-At}\dot{\bm{x}}(t) - e^{-At}A\bm{x}(t) = e^{-At}B\bm{u}(t).
\end{align*}
Next, we see that the left-hand side is the result of a product rule where $\frac{d}{dt}(e^{-At}\bm{x}(t)) = e^{-At}\dot{\bm{x}}(t) - Ae^{-At}\bm{x}(t)$, recalling that functions of matrices can switch orders of multiplication, such that $Ae^{-At} = e^{-At}A$. Hence, we can write our equation as
\begin{align*}
\frac{d}{dt}(e^{-At}\bm{x}(t)) = e^{-At}B\bm{u}(t),
\end{align*}
and integrate both sides from $t = 0$ to $t = T$ to yield
\begin{align*}
e^{-AT}\bm{x}(T) - \bm{x}(0) = \int_0^T e^{-At}B\bm{u}(t)dt.
\end{align*}
We note the matrix exponential at $t = 0$ becomes $e^{-A0} = I$ from the Taylor series. Next, we move the initial state $\bm{x}(0)$ to the right hand side, and multiply by $e^{AT}$
\begin{align*}
e^{AT}e^{-AT}\bm{x}(T) = e^{AT}\bm{x}(0) + e^{AT}\int_0^T e^{-At}B\bm{u}(t)dt.
\end{align*}
Finally we use the fact that $e^{AT}$ and $e^{-AT}$ are inverses of each other where $e^{AT}e^{-AT} = I$, and we bring $e^{AT}$ into the integral to derive the system's response to control input
\begin{align}
\label{eq:control}
\bm{x}(T) = \underbrace{e^{AT}\bm{x}(0)}_{\mathrm{natural}} + \underbrace{\int_0^T e^{A(T-t)}B\bm{u}(t)dt}_{\mathrm{controlled}}.
\end{align}
Intuitively, we see that the first part of the response, $e^{AT}\bm{x}(0)$, is just the natural evolution of our system from an initial state, and that the second part of the response is a convolution of our mapped inputs, $B\bm{u}(t)$, with the impulse response. We will next take advantage of the convolution's property of linearity to draw powerful relations between the state evolution, control input, and system structure.

\subsection{Linear Relation Between the Convolution and Control Input}
\label{subsec:7} 
Previously, we focused on the evolution of a neural system in response to a known control input $\bm{u}(t)$ in Eq.~\ref{eq:control}. However, our goal is to design a control input that drives our neural system to some desired final state that may stabilize an epileptic seizure \cite{taylor2015optimal}, or aid in memory recall \cite{ezzyat}. In this scenario, we fix the initial state $\bm{x}(0) = \bm{x}_0$ and the final state $\bm{x}(T) = \bm{x}_T$ as constants, and we search for an input $\bm{u}(t)$ that satisfies
\begin{align*}
\int_0^T e^{A(T-t)}B\underbrace{\bm{u}(t)}_{\textrm{variable}}dt = \underbrace{\bm{x}(T) - e^{AT}\bm{x}(0)}_{\textrm{constant}}.
\end{align*}
This formulation is a linear equation with a structure that is similar to a typical system of linear equations used in regression, $M\bm{v} = \bm{b}$, where $\bm{v}$ is the variable, $\bm{b}$ is a constant vector, and matrix $M$ is the linear function acting on $\bm{v}$. Here, the control input $\bm{u}(t)$ is the variable, $\bm{x}(T)-e^{AT}\bm{x}(0)$ is the constant vector, and the convolution 
\begin{align*}
\mc{L}(\bm{u}(t)) = \int_0^T e^{A(T-t)}B\bm{u}(t)dt,
\end{align*} 
is the linear function acting on our control inputs. By linear function, we mean that for two control inputs $\bm{u}_1(t), \bm{u}_2(t)$, if $\mc{L}(\bm{u}_1(t)) = \bm{c}_1$, and $\mc{L}(\bm{u}_2(t)) = \bm{c}_2$, then a weighted sum of inputs yields the same weighted sum of outputs, such that
\begin{align}
\label{eq:linearity}
\mc{L}(a\bm{u}_1(t) + b\bm{u}_2(t)) = a\bm{c}_1 + b\bm{c}_2.
\end{align}
This linearity allows us to treat solutions to our control function problem the same as solutions to our linear system of equations. Specifically, suppose control input $\bm{u}^*(t)$ was a \emph{particular solution} to our control problem such that $\mc{L}(\bm{u}^*(t)) = \bm{x}_T - e^{AT}\bm{x}_0$, and $\bm{u}_1(t), \bm{u}_2(t), \dotsm$ were \emph{homogeneous solutions} such that $\mc{L}(\bm{u}_i(t)) = \bm{0}$. Then the set of all valid control inputs is given by $\bm{u}(t) = \bm{u}^*(t) + \sum_i a_i\bm{u}_i(t)$, because
\begin{align*}
\mc{L}(\bm{u}(t)) &= \mc{L}(\bm{u}^*(t)) + \sum_i \mc{L}(a_i\bm{u}_i(t))\\
                  &= \bm{x}_T - e^{AT}\bm{x}_0 + \sum_i a_i\bm{0}\\
                  &= \bm{x}_T - e^{AT}\bm{x}_0.
\end{align*}

\subsection{Controllability}
For any system, we would first like to know if a particular solution exists to the control problem described above. A system is \emph{controllable} if there is a control input that brings our system from any initial state to any final state in finite time. For nonlinear systems, if we know that the input $\bm{u}^*(t)$ brings our system from the initial state $\bm{0}$ to some final state $\bm{x}_T$, there is in general no way to know what input will take our system to a scaled final state $a\bm{x}_T$. 

In contrast, due to the linearity of our convolution operator, we know that a scaled input $a\bm{u}^*(t)$ will produce a scaled output $\mc{L}(a\bm{u}^*(t)) = a\bm{x}_T$. Further, any $N$-dimensional vector can be written as a weighted sum of $N$ \emph{linearly independent} vectors $\bm{v}_1, \bm{v}_2, \dotsm, \bm{v}_N$. Here, linear independence means that no vector $\bm{v}_i$ in the set can be written as a weighted sum of the remaining vectors $\bm{v}_{j\neq i}$. For example, a column vector $\bm{a} = [a_1; a_2; \dotsm; a_N]$ can be written as the weighted sum
\begin{align*}
\underbrace{\begin{bmatrix}
a_1\\ a_2\\ \vdots\\ a_N
\end{bmatrix}}_{\bm{a}}
=
a_1 
\underbrace{\begin{bmatrix}
1\\0\\\vdots\\0
\end{bmatrix}}_{\bm{v}_1}
+
a_2 
\underbrace{\begin{bmatrix}
0\\1\\\vdots\\0\\
\end{bmatrix}}_{\bm{v}_2}
+
\dotsm
+
a_N 
\underbrace{\begin{bmatrix}
0\\0\\\vdots\\1
\end{bmatrix}}_{\bm{v}_N},
\end{align*}
where none of the vectors $\bm{v}_i$ can be written as a weighted sum of remaining vectors $\bm{v}_{j\neq i}$. Hence, our system is controllable if we can find input functions $\bm{u}_1(t), \dotsm, \bm{u}_N(t)$ that reach $N$ linearly independent vectors $\mc{L}(\bm{u}_1(t)), \dotsm, \mc{L}(\bm{u}_N(t))$, because then we can always reach any final state from any initial state through the weighted sum
\begin{align*}
\underbrace{\bm{x}_T - e^{AT}\bm{x}_0}_{\bm{a}} = 
a_1 \underbrace{\mc L(\bm{u}_1(t))}_{\bm{v}_1} + 
a_2 \underbrace{\mc L(\bm{u}_2(t))}_{\bm{v}_2} +
\dotsm + 
a_N \underbrace{\mc L(\bm{u}_N(t))}_{\bm{v}_N},
\end{align*}
through the control input $\bm{u}(t) = a_1\bm{u}_1(t) + a_2\bm{u}_2(t) + \dotsm + a_N\bm{u}_N(t)$. This information of reachable states is encoded in the \emph{controllability matrix}
\begin{align}
\label{eq:controllability}
\mc{C} = 
\begin{bmatrix}
B, & AB, & A^2B, & \dotsm, & A^{N-1}B
\end{bmatrix},
\end{align}
where the \emph{rank} of this matrix (given by the number of linearly independent columns of $\mc C$) tells us how many of these $N$ independent vectors can be reached using control input. If this rank $= N$, then the system is controllable and can reach all states. Further, if some vector $\bm{x}_T - e^{AT}\bm{x}_0$ can be written as a weighted sum of the columns of $\mc C$, then there exists a control input that drives the system from $\bm{x}_0$ to $\bm{x}_T$. This set of vectors spanned by the columns of $\mc C$ is called the \emph{controllable subspace}.

As an example in our two unit system, the controllability matrix is written as
\begin{align*}
\mc C = \begin{bmatrix}
1 & -1\\
0 & 0
\end{bmatrix},
\end{align*}
which is \emph{not} controllable, because the rank of $\mc C$ is 1. To consider the controllable subspace, notice that the columns of $\mc C$ only have non-zero entry in the first row. Hence, the controllable subspace contains any desired value of $x_1(T)$, but excludes all values of $x_2(T)$. Intuitively, this loss of controllability arises because $x_2$ does not receive an input, nor is it affected by $x_1$. Hence, there is no way to influence the activity of $x_2$ in a desired way.

\subsection{Minimum Energy Control}
Once we know a system is controllable, we would like to determine the control input function $\bm{u}(t)$ that transitions our system from initial $\bm{x}_0$ to final $\bm{x}_T$ states. However, there are often limitations on the input magnitude such as electrical and thermal damage of neural tissue, or battery life of chronic implanted stimulators. Due to the system's linearity, we can not only find an input function, but an optimal one $\bm{u}^*(t)$ that minimizes input cost.

First, we must define a measure of the \emph{size} of our control input functions $\bm{u}(t)$. In many applications of electrical and electromagnetic stimulation, the cost of control scales quadratically with the input such as resistive heating with electrical current. This quadratic measure of size is mathematically and intuitively defined using the \emph{inner product}. For $N$-dimensional column vectors of numbers, $\bm{a}$, the inner product is the well known \emph{dot product}
\begin{align*}
<\bm{a},\bm{a}> = a_1^2 + a_2^2 + \dotsm + a_N^2 = \bm{a}'\bm{a},
\end{align*}
where $\bm{a}'$ is the transpose that turns column vector $\bm{a}$ into a row vector. We see that doubling $\bm{a}$ will quadruple the inner product. For $k$-dimensional column vectors of functions, $\bm{a}(t)$, the inner product is similarly defined as
\begin{align*}
<\bm{a}(t),\bm{a}(t)> = \int a_1^2(t) + a_2^2(t) + \dotsm + a_N^2(t)dt = \int \bm{a}'(t)\bm{a}(t)dt,
\end{align*}
that has the same quadratic relation. Hence, we define the \emph{control energy} as
\begin{align}
\label{eq:energy}
E = <\bm{u}(t),\bm{u}(t)>.
\end{align}

Now that we have a measure of how large an input is, we wish to find a minimal input $\bm{u}^*(t)$ that minimizes the control energy. This task is analogous to a typical linear system of equations, $M\bm{v} = \bm{b}$, where we want to find $\bm{v}^*$ that solves the equation with the smallest cost $<\bm{v}^*,\bm{v}^*>$. Here, if $M$ has full row rank where the rows of $M$ are linearly independent, then the minimum solution is given by the equation for least squares $\bm{v}^* = M'(MM')^{-1}\bm{b}$. Here, $M'$ is the transpose, or \emph{adjoint} of $M$.

This same principle holds for our linear system $\mc L(\bm{u}(t)) = \bm{x}_T - e^{AT}\bm{x}_0$, where we want to find $\bm{u}^*(t)$ that solves the equation with the smallest cost $<\bm{u}^*(t),\bm{u}^*(t)>$. However, while matrix $M$ inputs a vector of numbers $\bm{v}$ and outputs a vector of numbers $\bm{b}$, our linear function $\mc L$ inputs a vector of functions and outputs a vector of numbers. Hence, we need to carefully define the transpose, or adjoint $\mc L'$. In the case of matrix $M$, the adjoint preserves the inner product between inputs and outputs such that
\begin{align*}
<M\bm{v},\bm{b}> &= <\bm{v},M'\bm{b}>\\ (M\bm{v})'\bm{b} &= \bm{v}' (M'\bm{b}).
\end{align*}
Identically, for state transition $\bm{x} = e^{AT}\bm{x}_0 - \bm{x}_T$, the adjoint of $\mc L$ preserves the inner product between the vectors of input functions $\bm{u}(t)$, and output numbers $\bm{x}$ as
\begin{align*}
<\mc L(\bm{u}(t)), \bm{x}> &= <\bm{u}(t), \mc L'(\bm{x})>\\
\left(\int_0^T e^{A(T-t)}B\bm{u}(t)dt\right)' \bm{x} &= \int_0^T \bm{u}'(t)(B'e^{A'(T-t)}\bm{x})dt.
\end{align*}
Notice that the inner product on the left is over vectors of numbers, while the inner product on the right is over vectors of functions. Then, we see that our adjoint is
\begin{align*}
\mc L'(\bm{x}) = B' e^{A'(T-t)}\bm{x},
\end{align*}
and takes as input a vector of numbers, and outputs a vector of functions. Then, just as our system $M\bm{v} = \bm{b}$, the minimum input $\bm{u}^*(t)$ is given by 
\begin{align}
\label{eq:minimum_trajectory}
\bm{u}^*(t) = \mc L'(\mc L \mc L')^{-1}(\bm{x}_T - e^{AT}\bm{x}_0).
\end{align}
Finally, through substitution into Eq.~\ref{eq:energy}, we can write the minimum control energy as
\begin{align}
\label{eq:minimum_energy}
E_{\mathrm{min}} = (\bm{x}_T-e^{AT}\bm{x}_0)' (\mc L \mc L')^{-1} (\bm{x}_T-e^{AT}\bm{x}_0).
\end{align}

In conclusion, we point out the crucially important term of the minimum energy, $\mc L \mc L'$, as the \emph{controllability Gramian} written as
\begin{align}
\label{eq:gramian}
W_c(T) = \mc L \mc L' = \int_0^T e^{A(T-t)}BB'e^{A'(T-t)}dt.
\end{align}
First, we notice that this Gramian is only a function of the underlying neural relationships, $A$, the matrix determining where the inputs are placed, $B$, and time $T$. Next, we notice that $W_c(T)$ is actually an $N \times N$ matrix, and can therefore be numerically evaluated and analytically studied. Finally, we see that if our system begins at an initial state of $\bm{x}_0 = \bm{0}$, then the minimum energy can be written
\begin{align*}
E_{\mathrm{min}} = \bm{x}_T' W_c^{-1}(T) \bm{x}_T,
\end{align*}
where the role of neural interactions and stimulation parameters on our ability to control the system is fully encapsulated in the Gramian. This ability to decouple the states $\bm{x}_T$ from the neural interactions and stimulation parameters $A,B,T$ is a powerful tool for studying and designing control properties of neural systems.

\section{Mapping Network Architecture to Control Properties}
\label{sec:4}
By formulating our neural system in a linear way, we can solve difficult problems such as predicting the system's response to control, finding the set of states that the system can reach, and designing efficient input stimuli, without the need to try every control input and simulate every trajectory. Further, by directly mapping control properties to neural activity and network architecture in an algebraic way, we can study how features of interaction patterns impact our ability to control neural activity \cite{tang2018control}. As an active area of research, the variety of questions being asked and systems being studied is very large, and require simultaneous innovations in experiment, computation, and theory. In this section, we will describe a few recent applications and advances.

\subsection{Neuronal Control in Model Organisms}
While most neural systems are too large to empirically measure activity and connectivity or to analyze numerically, there do exist a few sufficiently simple model organisms. Among these is the worm \emph{Caenorhabditis elegans} \cite{white} with several hundred neurons that can be recorded from simultaneously \cite{nguyen}. Even for such a small system, it is difficult to map the functional form of how activity in neuron $i$ affects the activity in neuron $j$. However, the presence or absence of connections between neurons in this organism, and by consequence the presence or absence of elements in the connectivity matrix $A$, is well known. 

Advances in the study of \emph{structural controllability} \cite{lin} allow us to ask questions about our ability to control a system given only the binary presence or absence of edges. Colloquially, this framework focuses on connectivity matrices $A$ where non-zero entries can only exist in the presence of binary edges, and can be used to determine whether the system is controllable for \emph{most} values where an edge is present. Using this framework, recent work has sought to determine whether the removal of certain neurons in \emph{C. elegans} will reduce structural controllability \cite{yan}. Specifically, the modeling involves input to the sensory receptor neurons as the control input that is mapped to the system through a matrix $B$, and the connectivity between neurons and muscle cells through a matrix $A$. Further, instead of recording the activity of each neuron, the motion of muscles was recorded. This framework involves the appended control framework
\begin{align*}
\dot{\bm{x}}(t) &= A\bm{x}(t) + B\bm{u}(t)\\
\bm{y}(t) &= C\bm{x}(t),
\end{align*}
where $\bm{y}(t)$ represents the states (muscles) that are measured, and $C$ is the map from neurons and muscles $\bm{x}(t)$ to the measured output \cite{towlson}. Here, the authors find that the ablation of a neuron not previously implicated in motion, PDB, decreased structural controllability, significantly reducing ventral bias in deep body bends in \emph{C. elegans}. 

\subsection{State Transitions in the Human Brain}
While neuron-level structural synapses map most directly to functional relationships between neurons, there are also well-characterized structural connections between larger-scale brain regions. These connections contain thick bundles of myelinated axonal fibers that run throughout the brain, and are thought to play a crucial role in coupling the activity of distant brain regions \cite{avena2017communication}. These fibers are resolved by measuring water diffusion throughout the brain using magnetic resonance \cite{taylor}, and tracing fibers along this diffusion field using computational algorithms \cite{basser}. The whole brain is typically divided into hundreds to thousands of discrete brain regions using a variety of parcellation schemes \cite{hagmann,power}, and the strength of fibers between these regions comprise the connectivity matrix $A$ \cite{bassett2018nature}.

Such region-level study of brain dynamics has led to the discovery of macroscopic functional organization in the human brain at rest \cite{raichle} and during various cognitively demanding tasks \cite{sporns2016modular}. Here, brain activity can be empirically measured through methods such as magnetic resonance imaging (blood oxygen level dependent) or electrophysiology (aggregate electrical activity). Of particular interest are large-scale functional brain networks that display stereotyped changes in activity patterns during tasks that demand certain cognitive or sensorimotor processes \cite{bressler}. Here, it is thought that the brain uses underlying structural connections to support circuit-level coordination, as well as to guide itself to specific patterns of activity using \emph{cognitive control} \cite{gu2015controllability,medaglia2018functional}.

Recent work has begun formulating cognitive control as a linear systems problem \cite{gu2015controllability,betzel,tang2017developmental,gu2017optimal,cornblath2018sex}, where matrix $A$ is the network of white matter connections between brain regions, $B$ represents the regions that were chosen to be responsible for control, and $\bm{x}(t)$ represents the activity of each region over time. Specifically in \cite{betzel,gu2017optimal}, the authors quantify cognitive states as vectors corresponding to activity in the brain regions during cognitive tasks, and compute the minimum control energy Eq.~\ref{eq:minimum_trajectory} to transition between cognitive states for various sets of control regions. Colloquially, if a set of regions requires less input energy to transition between cognitive states, then those regions may easily transition the whole brain between these states along an optimal trajectory given they are responsible for cognitive control. Moreover, individual differences in the minimal control energy are correlated with individual differences in performance on cognitive control tasks \cite{cui2018optimization}. In complementary studies, individual differences in controllability statistics calculated for distinct regions of the brain are correlated with individual differences in measures of cognitive control assessed with common neuropsychological test batteries \cite{tang2017developmental,cornblath2018sex}.

\section{Methodological Considerations and Limitations}
\label{sec:5}
While the theory of linear systems is a powerful quantitative framework for studying and controlling dynamical neural systems, there are several important caveats. Here we mention three: dimensionality and numerical stability, model validation and experimental data, and the assumption of linearity.

\subsection{Dimensionality and Numerical Stability}
The benefit of studying linear systems is that we take difficult and largely intractable questions of controllability and control input design, and greatly simplify them into algebraic problems of computing objects like the controllability matrix Eq.~\ref{eq:controllability} and the controllability Gramian Eq.~\ref{eq:gramian}. However, these matrices scale quadratically with the number of neural units, and numerical calculations and manipulations using these matrices quickly face computational issues.

Most viable approaches to dealing with these issues involve numerically representing the elements of our matrices, and performing algebraic operations. However, these representations are imperfect, as it is impossible to completely represent irrational numbers such as $\pi$. Hence, the matrices are truncated to \emph{numerical precision}, and this truncation error propagates with each computation. Further, the propagation of error tends to scale faster than the number of dimensions. This issue is prevalent in the computation of the state-transition matrix \cite{moler}, as well as in the calculation of the controllability Gramian and its inverse. With the application of this theory to high dimensional neural systems, the study of useful controllability metrics is an active area of research \cite{pasqualetti}.

\subsection{Model Validation and Experimental Data}
A fundamental limitation for modeling any neural system is the ability to empirically and accurately measure model parameters and variables. A crucial parameter is the network of connectivity encoded by our adjacency matrix $A$, where the element in the $i$-th column and $j$-th row models the effect of unit $i$ on the rate of change of unit $j$. While we typically use the structural connections in synapses between neurons, or bundles of axons between brain regions as a proxy for $A$, it is very difficult to measure the true functional effect that activity in unit $i$ has on activity in unit $j$, particularly for large systems. This problem is exacerbated by further methodological limitations such as the inability to resolve directionality of connections in diffusion tractography. Along these lines, many statistical and autoregressive methods have been developed to infer functional relationships from recordings of neural activity \cite{granger,seth2015granger,barnett2018misunderstandings,friston2011functional,mcintosh2012tracing}, and to use that inferred activity to better understand control \cite{becker2015large}. However, the degree of causality in these methods as measured by true response to external stimuli remains controversial.

Another such fundamental limitation is our inability to fully measure every state of the system. The state-space representation of our model requires that every state is observed. However, it is impossible to simultaneously record the activity of every neuron in almost all biological systems, although this recording has been achieved in sufficiently simple organisms \cite{nguyen}. As a result of only being able to observe a small subset of the full state-space, these models of interactions may become largely descriptive and phenomenological in nature. In response, there is a continuing effort to improve the spatial and temporal resolution of neuroimaging methods \cite{stosiek}.

\subsection{Assumption of Linearity}
An inherent limitation is the lack of generality in our linear approximation of the full nonlinear neural dynamics. In response, there is a sizable quantity of research studying the control properties of nonlinear dynamical systems \cite{motter2015networkcontrology}. An interesting bridge between these two disciplines exists in the theory of the Koopman or composition operator \cite{koopman}. The underlying benefit of this theory is that, while our system of equations may evolve nonlinearly in time given the current set of $N$ states, there may exist a higher-dimensional set of $M > N$ state variables in which the dynamical system does evolve linearly \cite{brunton}. While the extension of linear systems theory to actually controlling this higher-dimensional system may be limited, it remains a promising future area of research.

\section{Open Frontiers}
\label{sec:6}

Many exciting and open frontiers exist in the study of brain network dynamics using linear systems theory. Here we constrain our remarks to three main topic areas, but freely admit that this discussion is far from comprehensive. First, we describe opportunities in the further development of useful controllability statistics as well as in the development of foundational theory linking control profiles to the system's underlying network architecture. Second, we underscore the need for a better understanding of how control is implemented in the brain, how control strategies might depend on context, and how control processes could facilitate the effective manipulation of information. Third, we describe the relevance of the modeling efforts we discussed here for our understanding of neurological disease and psychiatric disorders as well as the development of personalized and targeted therapeautic interventions for alterations in mental health. 

\subsection{Theory and Statistics}
Linear systems theory has its basis in a rich literature stemming from now well-developed areas of mathematics, physics, and engineering \cite{kailath1980linear}. Yet, much is still unknown about exactly how the network topology of a given unit-to-unit interaction pattern impacts the capacity for control, the trajectories accessible to the systems, and the minimum control energy. Some preliminary efforts have begun to make headway by using linear network control theory to derive accurate closed-form expressions that relate the connectivity of a subset of structural connections (those linking driver nodes to non-driver nodes) to the minimum energy required to control networked systems \cite{kim2018role}. Further work is needed gain an intuition for the role of higher order structures (e.g., cycles) in the control of the networked system, and any dependence on edge directionality \cite{xiao2015effects}. Moreover, it would be fruitful in the future to further develop a broader set of controllablity statistics, extending beyond node controllability \cite{pasqualetti}, and edge controllability \cite{pang2017universal}, to the control of motifs \cite{whalen2015observability}. Finally, throughout such investigations it will be useful to understand which features of control are shared across networks with various topologies, versus those features which are specific to networks with a particular topology \cite{wuyan2018benchmarking,tu2018warnings,menara2017structural}.

\subsection{Context, Computations, and Information Processing}

Despite the emerging appreciation that linear systems theory has considerable utility in the study of cognitive function, we still know very little about exactly how control is implemented in the brain, across spatial scales, and capitalizing on the unit-to-unit interaction patterns at each of those scales.  Some initial evidence suggests that features of synaptic connectivity -- and particularly autaptic connections -- can serve to tune the excitability of the neural circuit, altering its controllability profile and propensity to display synchronous bursts of activity \cite{wiles2017autaptic}. Complementary evidence also at the cellular scale demonstrates how intrinsic network structure and exogeneous stimulus patterns together determine the manner in which a stimulus propagates through the network, with important implications for cognitive faculties that require persistent activation of neuronal patterns such as working memory and attention \cite{ju2018network}. There are interesting similarities between these observations and evidence at larger spatial scales, which suggests that the architecture of white matter tracts connecting brain areas can be used to infer the probability with which the brain persists in certain states \cite{cornblath2018context}. Such conceptual similarities motivate concerted efforts to better understand how the architecture of brain networks across spatial scales supports information processing and cognitive computations, and how those processes and computations might depend on the context in which the brain is placed. Formally, it would be interesting to consider context as a form of exogeneous input to the system, in a manner reminiscent of how we currently consider brain stimulation \cite{tang2018control}. We speculate that such a formulation of the problem could help to explain a range of observations, such as the ability of cognitive effort to suppress epileptic activity \cite{muldoon2018locally}.

\subsection{Disease and Intervention}

The fact that controllability can depend on network topology \cite{wuyan2018benchmarking,kim2018role} and can be altered by edge pruning \cite{mengiste2015effect}, suggests that it might also be a useful biomarker in some neurological diseases and psychiatric disorders, many of which are associated with changes in the structural topology of neural circuitry at various spatial scales \cite{braun2018maps,stam2014modern}. Indeed, recent studies have reported differences in controllability statistics estimated in brain networks of patients with bipolar disorder \cite{jeganathan2018fronto}, temporal lobe epilepsy \cite{bernhart2018hippocampal}, and mild traumatic brain injury \cite{gu2017optimal}. In a complementary line of work, studies are beginning to ask whether the altered controllability profiles of brain networks in these patients could help to inform the development of more targeted interventions for their illness, in the form of brain stimulation \cite{muldoon2016stimulation,taylor2015optimal}, pharmacological agents, or cognitive behavioral therapy. Other efforts have begun to consider symptoms of a given disease as a network, and to identify symptoms predicted to have high impulse response in the patient's daily life \cite{yang2018socioemotional}. It would be interesting in future to determine whether the linear systems approach could be useful in more carefully formalizing that problem as a network control problem, which in turn could be used to determine which symptom to treat in order to move the entire symptom network towards a healthier state \cite{lydonstaley2018digital}.

\section{Acknowledgements}
\noindent We gratefully acknowledge comments and feedback from Arian Ashourvan, Ann E. Sizemore, Melody X. Lim, Jennifer A. Stiso, Erin G. Teich, Teresa Karrer, Zhixin Lu, Harang Ju, and Eli J. Cornblath. We also thank Ann E. Sizemore for generous assistance with and input on schematic figure construction. JZK acknowledges support from the NIH T32-EB020087, PD: Felix W. Wehrli, and the National Science Foundation Graduate Research Fellowship No. DGE-1321851. DSB acknowledges support from the John D. and Catherine T. MacArthur Foundation, the Alfred P. Sloan Foundation, the Paul G. Allen Foundation, the Army Research Laboratory through contract number W911NF-10-2-0022, the Army Research Office through contract numbers W911NF-14-1-0679 and W911NF-16-1-0474, the National Institute of Health (2-R01-DC-009209-11, 1R01HD086888-01, R01-MH107235, R01-MH107703, R01MH109520, 1R01NS099348 and R21-M MH-106799), the Office of Naval Research, and the National Science Foundation (BCS-1441502, CAREER PHY-1554488, BCS-1631550, and CNS-1626008).The content is solely the responsibility of the authors and does not necessarily represent the official views of any of the funding agencies.

\newpage
\section{Problems}
\label{sec:7}
\begin{itemize}
	\item \textbf{Problem 1:} Linearize the following system about point $x_1^* = 1, x_2^* = -1, x_3^* = 0$,
	\begin{align*}
	\begin{bmatrix}
	\dot{x}_1(t)\\
	\dot{x}_2(t)\\
	\dot{x}_3(t)
	\end{bmatrix}
	=
	\begin{bmatrix}
	-x_1^2(t) - 2x_2(t) + x_3(t) - 1\\
	2x_1(t) - 2x_2^2(t) + 2x_3(t)\\
	x_1(t)x_2(t) - x_3(t) + 1
	\end{bmatrix}.
	\end{align*}
	and demonstrate that this point is a fixed point where $\dot{x}_1 = \dot{x}_2 = \dot{x}_3 = 0$.
	\\~\\
	
	\item \textbf{Problem 2:} Prove that the matrix exponential of $A = \begin{bmatrix}a & 0\\ 0 & b\end{bmatrix}$ is
	\begin{align*}
	e^{A}
	=
	\begin{bmatrix}
	e^a & 0\\
	0 & e^b
	\end{bmatrix},
	\end{align*}
	using the Taylor series of the scalar and matrix exponentials.
	\\~\\
	
	\item \textbf{Problem 3:} Prove that the system response to control 
	\begin{align*}
	\bm{x}(t) = e^{At}\bm{x}(0) + \int_0^t e^{A(t-\tau)}B\bm{u}(\tau)d\tau,
	\end{align*} 
	satisfies the dynamical equation $\dot{\bm{x}}(t) = A\bm{x}(t) + B\bm{u}(t)$ by substitution.
	\\~\\
	
	\item \textbf{Problem 4:} Prove that the convolution operator
	\begin{align*}
	\mc L(\bm{u}(t)) = \int_0^Te^{A(T-\tau)}B\bm{u}(\tau)d\tau
	\end{align*}
	is linear according to Eq.~\ref{eq:linearity}; that is, if $\mc L(\bm{u}_1(t)) = \bm{c}_1$, and $\mc L(\bm{u}_2(t)) = \bm{c}_2$, then demonstrate that $\mc L(a\bm{u}_1(t) + b\bm{u}_2(t)) = a\bm{c}_1 + b\bm{c}_2$. 
	\\~\\
	
	\item \textbf{Problem 5:} Determine if the following system is controllable
	\begin{align*}
	\begin{bmatrix}
	\dot{x}_1(t)\\
	\dot{x}_2(t)\\
	\dot{x}_3(t)
	\end{bmatrix}
	=
	\begin{bmatrix}
	0 & 1 & 0\\
	0 & 0 & 1\\
	1 & 0 & 0
	\end{bmatrix}
	\begin{bmatrix}
	x_1(t)\\ x_2(t)\\ x_3(t)
	\end{bmatrix}
	+
	\begin{bmatrix}
	1 \\ 0 \\ 0
	\end{bmatrix}
	u(t),
	\end{align*}
	by constructing the controllability matrix.
	\\~\\
	
	\item \textbf{Problem 6:} Determine for what value of $a$ the system is not controllable
	\begin{align*}
	\begin{bmatrix}
	\dot{x}_1(t)\\
	\dot{x}_2(t)\\
	\dot{x}_3(t)
	\end{bmatrix}
	=
	\begin{bmatrix}
	0 & 0 & 0\\
	1 & 1 & 0\\
	1 & 0 & a
	\end{bmatrix}
	\begin{bmatrix}
	x_1(t)\\ x_2(t)\\ x_3(t)
	\end{bmatrix}
	+
	\begin{bmatrix}
	1 \\ 0 \\ 0
	\end{bmatrix}
	u(t),
	\end{align*}
	by constructing the controllability matrix.
	\\~\\
	
	\item \textbf{Problem 7:} Derive the minimum energy equation Eq.~\ref{eq:minimum_energy} 
	\begin{align*}
	E_{\mathrm{min}} = (\bm{x}_T-e^{AT}\bm{x}_0)' (\mc L \mc L')^{-1} (\bm{x}_T-e^{AT}\bm{x}_0),
	\end{align*}
	by substituting the minimum input $\bm{u}^*(t)$ into the control energy Eq.~\ref{eq:energy}
	\begin{align*}
	E = <\bm{u}(t),\bm{u}(t)>.
	\end{align*}
	\\~\\
	
	\item \textbf{Problem 8:} Show that the controllability Gramian can be written
	\begin{align*}
	W_C(T) = \int_0^T e^{A(T-t)}BB^Te^{A^T(T-t)}dt = \int_0^T e^{A\tau}BB^Te^{A^T\tau}d\tau,
	\end{align*}
	using the substitution $\tau = T - t$.
	\\~\\
	
	\item \textbf{Problem 9:} Show that the controllability Gramian for system 
	\begin{align*}
	A = \begin{bmatrix} a & 0\\ 0 & b \end{bmatrix}, \hspace{1cm} B = \begin{bmatrix}  1 & 0\\ 0 & 1 \end{bmatrix}
	\end{align*}
	is
	\begin{align*}
	W_C(T) = 
	\begin{bmatrix}
	\frac{1}{2a}\left(e^{2aT}-1\right) & 0\\
	0 & \frac{1}{2b}\left(e^{2bT}-1\right)
	\end{bmatrix}
	\end{align*}
	\\~\\
	
	\item \textbf{Problem 10:} Compute the minimum energy required for the system
	\begin{align*}
	A = \begin{bmatrix}
	\frac{1}{2} & 0\\
	0 & 2
	\end{bmatrix},
	\hspace{1cm}
	B = \begin{bmatrix}
	1 & 0\\
	0 & 1
	\end{bmatrix},
	\end{align*}
	to transition from initial state $\bm{x}(0) = \begin{bmatrix}0 \\ 0\end{bmatrix}$ to final state $\bm{x}(T) = \begin{bmatrix}1 \\ 2\end{bmatrix}$ in time $T = 1$.

\end{itemize}

%
%
%

\end{document}